\documentclass[prl,twocolumn,nofootinbib]{revtex4-1}
\usepackage{amsmath,amssymb}
\usepackage{graphicx}
\usepackage{slashed}

\newcommand{\be}{\begin{equation}}
\newcommand{\ee}{\end{equation}}

\begin{document} 

\title{Exact Renormalization of the Higgs Field}
\author{Kang-Sin Choi}
\email{kangsin@ewha.ac.kr}
\affiliation{Scranton Honors Program, Ewha Womans University, Seoul 03760, Korea}
\affiliation{Institute of Mathematical Sciences, Ewha Womans University, Seoul 03760, Korea}

\begin{abstract}
Using Wilsonian renormalization, we calculate the quantum correction to observable quantities, rather than the bare parameters, of the Higgs field. A physical parameter, such as a mass-squared or a quartic coupling, at an energy scale $\mu$ is obtained from that at a reference scale by integrating in the degrees of freedom in between. In this process, heavy modes decouple and the ultraviolet scale dependence is canceled in the observables. An exact renormalization group equation is parametrized by the low-energy scale $\mu$.
\end{abstract} 

\maketitle

Renormalization in quantum field theory informs us how to deal with the energy dependence of physical parameters.
In computing their quantum corrections including loops, divergences arise. Traditionally, quantum field theory aimed to remove these divergences; however, they indicate our limited knowledge of small-scale physics. These divergences have been brought under control through Wilsonian renormalization \cite{Wilson:1971bg,Polchinski:1983gv}. The associated ultraviolet cutoff is interpreted as the energy scale or the dimensionful parameter that signifies new physics. The dependence on the ultraviolet cutoff is further refined and understood as that of the infrared energy scale \cite{Wetterich:1992yh,Morris:1993qb,Morris:1998da}, revealing the direct connection to renormalization group equation \cite{Callan:1970yg,Symanzik:1970rt,Symanzik:1971vw}. We can follow how such parameters inherite to the low-energy ones. 

So far, the renormalization procedure has focused on the {\em bare} parameters of a given quantum field theory \cite{Choi:2023cqs}. In this letter, we perform Wilsonian renormalization for physically observable parameters, especially the Higgs mass. For instance, if we take a reference as the pole mass, its quantum correction is expressed as the difference of the bare parameters at different energy scales, canceling the cutoff dependence \cite{Choi:2023cqs}. 

Take a Higgs scalar $\phi(x)$ couped to a heavy scalar $X(x)$ and a fermion $t(x)$. We consider a Euclidian action
\be \begin{split} \label{Lhigh}
- {\cal L} &\supset \frac12 \partial_\mu \phi \partial^\mu \phi +\frac12 m_0^2 \phi^2 +  \frac{\lambda_0}{4} \phi^4  \\
& \quad + \frac12 \partial_\mu X \partial^\mu X + \frac12 M_0^2 X^2 + \frac{\kappa_0}4  \phi^2 X^2 \\
& \quad + \overline t_{L,R} \gamma^\mu \partial_\mu t_{L,R}+\frac{y_{t,0}}{\sqrt 2}  (v_0 + \phi) ( \overline t_L t_R + \overline t_R t_L).
\end{split}
\ee
Here, we expanded the Higgs field around the electroweak vacuum expectation value (VEV) $v_0$. Thus it is a real scalar field usually denoted by $h$. Other terms are omitted for simplicity.
We consider a hierarchy of energy scales
\be
   \Lambda  < M_0 < \Lambda_? ,
\ee
that is, we assume that the above Lagrangian is valid below an energy scale $\Lambda_?$. We shall be interested in the scale way below the physics of the field $X$.
In principle, we do not need $\Lambda$, but in practice, we do not know the necessary dimensionful parameter $M_0$. For $\phi$, we consider both cases $m_0 \ll M_0$ and $m_0 \sim M_0$.\footnote{We should deal with the renormalized masses, not the bare ones: see below. We are interested in the mass and the quartic coupling of the scalar field $\phi$, so except for those, we assume that the renormalized couplings are obtained in the same way. The corresponding quantities are denoted with the subscript as the scale.}

We calculate the mass of the field $\phi$ in low energy. 
Since the mass $M$ of the field $X$ is very large from the low-energy viewpoint, we expect it to be decoupled. We also separate, in the scalar field, the high-frequency part $\hat \phi(k)$, whose Euclidianized momentum is larger than $\mu$, as
\be
 \phi(k) \to \phi(k) + \hat \phi(k),
\ee
and let the remaining one $\phi(k)$ fluctuate below $\mu$ \cite{Wilson:1971bg}.
The relevant interactions are
\be \label{Masscorr}
 V = \frac{3\lambda_0}{2} \phi^2 \hat \phi^2 + \frac{\lambda_0}{4} \hat \phi^4+ \frac{\kappa_0}{4} \phi^2 X^2 + \frac{y_{t,0}}{\sqrt 2}  \phi ( \overline t_L t_R + \overline t_R t_L).
\ee

The effective mass operator is obtained by contracting two $\hat \phi$ fields in the interactions \eqref{Masscorr}, using the Feynman propagator \cite{Polchinski:1983gv}. We do not consider external momenta. For the high-frequency scalar, we have 
\be \begin{split}
 \Sigma_{\hat \phi}(\mu^2) &=  \frac32 \int_{\mu <|k|<\Lambda} \frac{d^4k}{(2\pi)^4} \frac{\lambda_k}{k^2+m_k^2} \\
&=  \frac32  \int_{\mu^2}^{\Lambda^{2}} \frac{dk^2k^2}{(4\pi)^2} \frac{\lambda_k}{k^2+m_k^2}. \\
\end{split}
\ee
Here, we took an arbitrary low energy scale $\mu < \Lambda$ and used the nontrivial propagator only in the range indicated in the integral. In Wilsonian renormalization, the cutoff $\Lambda$ is not the regularized infinity but the scale we specify. In the propagator, we used a renormalized mass $m_k$ {\em at} the energy scale $k$ and a quartic coupling $\lambda_k$ that we clarify shortly (see Eqs. (\ref{corrmass}) and (\ref{lambdarun})). This makes the definition of mass self-dependent, but we can approximate and calculate it perturbatively later.

Similarly, the heavy field $X$ and the top quark contribute to the mass by integrating out over the same region
\begin{align}
\tilde \Sigma_{X}(\mu^2) &= \frac{1}{4} 
 \int_{\mu^2}^{\Lambda^2} \frac{dk^2k^2}{(4\pi)^2} \frac{\kappa_k}{k^2+M_k^2}, \label{Xdecouple} \\
\tilde \Sigma_{t}(\mu^2) &=-\int_{\mu^2}^{\Lambda^2} \frac{dk^2k^2}{(4\pi)^2}\frac{y_t^2(k^2-m_{t,k}^2)}{(k^2+m_{t,k}^2)^2}.
\end{align}
In the low-energy theory, the mass is corrected as
\be \label{corrmass}
 m_\mu^2  \equiv m_0^2 + \sum \tilde \Sigma_i(\mu^2),
\ee
where the summation runs over all the contributions we consider.

Apparently, the mass correction depends on the high scale $M$ and the cutoff $\Lambda$ quadratically. We can further infer that all the bare parameters, including $m_0^2, \lambda_0 $ must also be dependent on the cutoff $ \Lambda_?$ and unknown physics beyond that. Therefore, the mass is not well-defined in low energy below $\Lambda$. This is the gauge hierarchy problem \cite{Gildener:1976ai}. Note that we do not address the problem of the smallness of $m_h^2$, but question the stability of the smallness against the correction by heavy fields \cite{Gildener:1976ai}. 

However, this has been about the {\em bare} parameters, which are not observables. Also, the cutoff $\Lambda$ and $\Lambda_?$ are not physical parameters but human-made and are to be matched by observables in the end \cite{Weinberg:1995mt}. In what follows, we show that we can mention {\em observable} mass and compute the quantum correction of it, whose result does not depend on high energy \cite{Choi:2023cqs}. 

Now we extract what {\em we can observe.} Only the combination (\ref{corrmass}) can be observable, so firstly we express it in reference to the pole mass.
The Feynman propagator for the low-energy scalar $\phi(x)$ with the momentum $k$ has a pole at the mass $k^2 = m_k^2$ defined in (\ref{corrmass}).
We define a ``pole mass'' $m_h$ as that satisfying
\be \label{polemass}
 m_h^2 = m_0^2 + \sum \tilde \Sigma_i(m_h^2).
\ee
This is a natural reference point, but any reference would be good. Since here we do not consider the kinematics of the field, the mass (\ref{corrmass}) is independent of $k$ and the field renormalization is not necessary. 

We can express the effective mass at scale $\mu$ in terms of the pole mass \cite{Choi:2023cqs}
\be \label{mcorrpole}
\begin{split} 
 m_\mu^2 &= m_h^2+ \sum \left[ \tilde \Sigma_i(\mu^2) -\tilde \Sigma_i(m_h^2) \right]  \\
 & \equiv  m_h^2+\delta m_h^2(\mu^2).
\end{split}
\ee
The unobservable bare mass $m_0$ is removed and the mass is understood as the quantum correction from the pole mass. Note that it depends on the scale $\mu$ we consider: like the upper limit $\Lambda$, we specified to which energy $\mu$ we run down. We consider quantum correction from this reference mass we can measure by experiment. This reference plays a similar role as the renormalization condition. The mass at different energy scales is now physical in the sense that we can measure it as well. 

Now, the mass correction consists of the combinations. For the high-frequency scalar $\hat \phi$, it is
\be \label{integratein}
 \tilde \Sigma_{\hat \phi}(\mu^2)- \tilde \Sigma_{\hat \phi}(m_h^2)  =  \frac{3}{2} \int^{m_h^2}_{\mu^2} \frac{dk^2 k^2}{(4\pi)^2}  \frac{\lambda_k}{k^2+m_k^2}.
\ee
This has a natural interpretation, faithfully following the Wilsonian program: The physical parameter (the mass squared) at scale $\mu^2$ is obtained from that at $m_h^2$ by integrating {\em in} the degrees of freedom from $m_h^2$ up to $\mu^2$. Considering multiple parameters, the renormalization group flow may branch, so we should technically understand this as the minus of the integrating out. Obviously, the integration is finite, as if we did not need any regularization. 

\begin{figure}[t!]
\begin{center}
\includegraphics[scale=0.44]{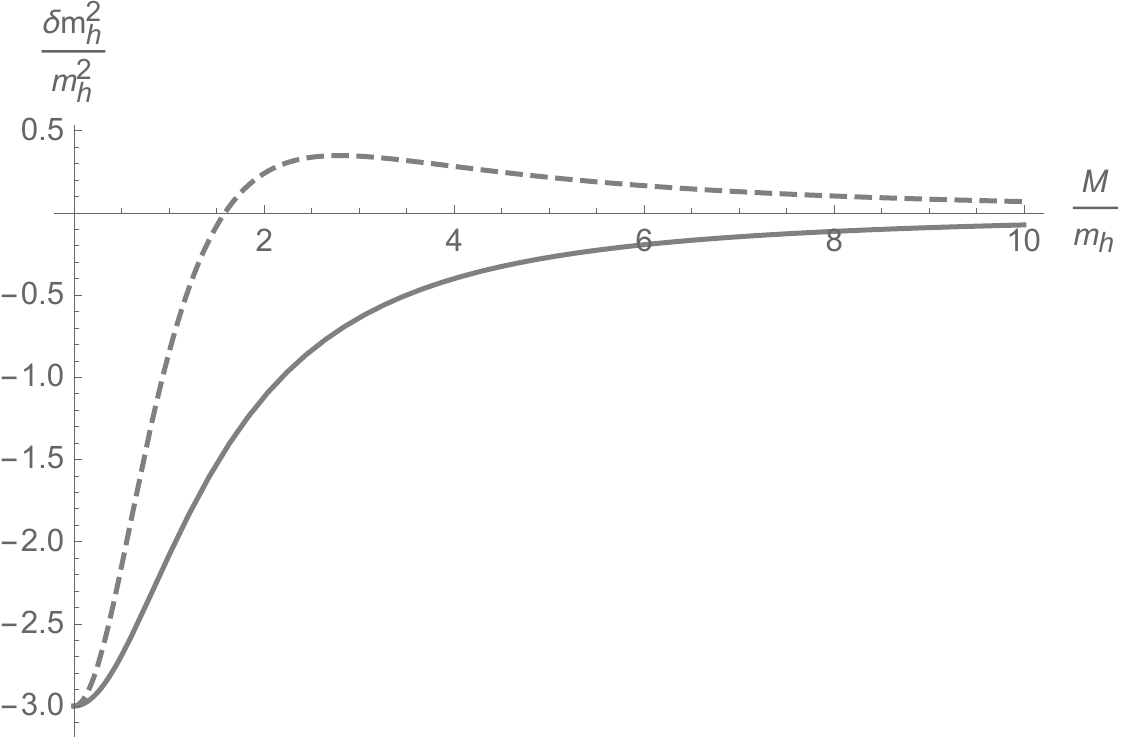}
\caption{Decoupling in the correction of a scalar mass squared $m_h^2$ by another scalar (fermion: dashed) with the mass $M$. This is a snapshot at $\mu = 2 m_h$. Its contribution starts from $1-\mu^2/m_h^2$ and quickly approaches zero.}
\label{f:corrscal}
\end{center}
\end{figure}

Since the running interval in the energy scale is short, we can approximate the mass to be constant around the Higgs pole mass. Also, we show shortly that the quartic coupling does not run much, so we approximate the dimensionless parameters as constants. Then, the mass correction (\ref{mcorrpole}) becomes 
\be \label{deltamfin}
\begin{split} 
  \delta m^2(\mu^2) &\simeq \frac{\kappa}{64\pi^2} \left[-\mu^2+m_h^2 -M^2 \log \frac{m_h^2+M^2}{\mu^2+M^2} \right]\\
  &+\frac{3\lambda }{32\pi^2}  \left[-\mu^2+m_h^2 -m_h^2 \log \frac{2m_h^2}{\mu^2+m_h^2} \right]\\
   &+ \frac{y_t^2}{16\pi^2} \bigg[- \mu^2 + m_h^2  - 3 m_t^2 \log \frac{\mu^2+m_t^2}{m_h^2+m_t^2}\\
   &\quad - 2m_t^4 \left. \left( \frac{1}{m_h^2+m_t^2} - \frac{1}{\mu^2 + m_t^2} \right) \right],
\end{split}
\ee
All the couplings are matched at $m_h$ and written down without subscripts.
The dependences on $\Lambda$, the quadratic and the logarithmic, are removed, as promised.  

We can show that all the terms from the same origin are of the same order. First, consider the correction from the heavy scalar $X$.
For large $M$, the logarithm is expanded as
$$
- \log \frac{m_h^2+M^2}{\mu^2+M^2} = \frac{\mu^2-m_h^2}{M^2} - \frac{\mu^4-m_h^4}{2M^4} + \frac{\mu^6-m_h^6}{3M^6} + \dots .
$$
Its dominant terms are canceled by the power-running part in $\mu$. 
Thus the first line in \eqref{deltamfin} becomes
\be
 \frac{\kappa}{64\pi^2} \left[   \frac{m_h^4-\mu^4}{2M^2} - \frac{m_h^6-\mu^6}{3M^4} + \dots \right]. 
\ee
This contribution actually becomes zero, contrary to na\"ive usual estimate, in the same way that the decoupling theorem applies \cite{Appelquist:1974tg}.
We draw the dependence of the scalar masses on the Higgs mass squared correction in Fig. \ref{f:corrscal}. 
Even if the scalar $X$ is light, its contribution is quickly suppressed for reasonable mass.

The same decoupling occurs for the heavy fermion $t$. For a large $m_t$, the last line in \eqref{deltamfin} is
\be 
  \frac{y_t^2}{16\pi^2}\sum_{n=1}^\infty (-1)^n \frac{2n-1}{n+1}\frac{m_h^{2n+2}-\mu^{2n+2}}{m_t^{2n}}.
\ee
For a considerably large $m_t$, the corresponding correction vanishes.\footnote{Here, we mean the case where the external heavy quark mass is given by some other mechanism than the electroweak Higgs. The top quark cannot be heavy because its mass is provided by the VEV multiplied by the order one Yukawa coupling.} Again, even for small mass, we see that heavy fermions also easily decouple as in Fig. \ref{f:corrscal}.

\begin{figure}[t]
\includegraphics[scale=0.45]{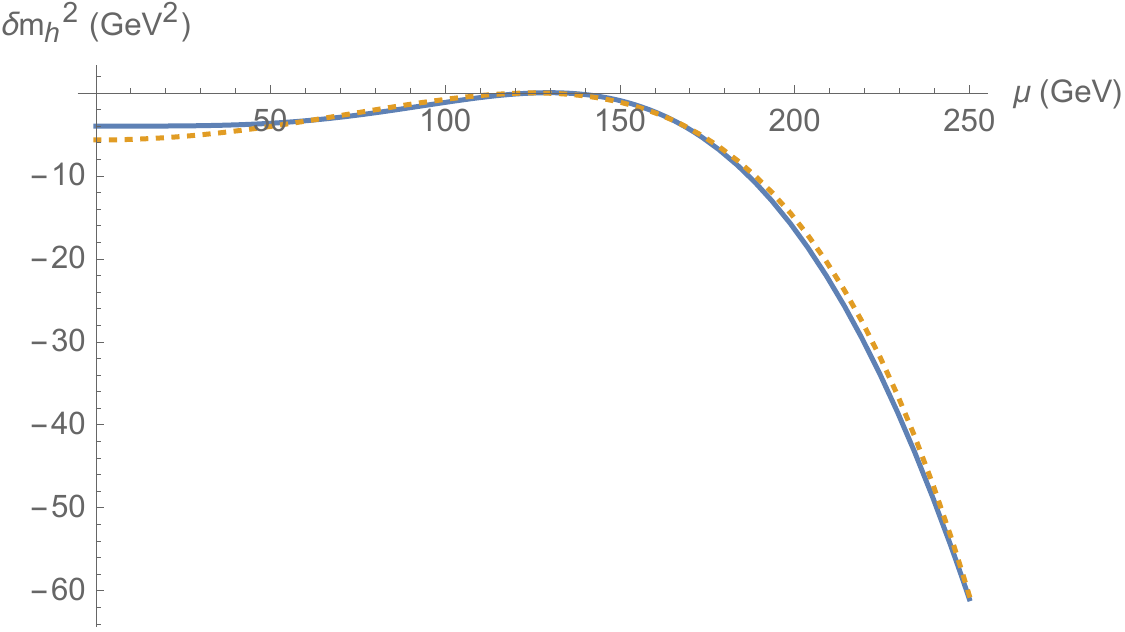}
\caption{The correction $\delta m_h^2$ to the Higgs pole mass squared $m_h^2$ at renormalization scale $\mu$, by integrating in the top quark loop and the scalar self-interacting loop (solid). It agrees with the conventional one-loop correction by top-quark (dotted) \cite{Choi:2023cqs}. We use $m_h = 125.4, m_t = 173$, all in GeV, and $y_t=1, \lambda =0.13$. The correction $\delta m_h^2$ is about $-0.39\%$ at 250 GeV.
\label{f:wilcorr}}
\end{figure}

The Higgs mass remains the same with good accuracy even if we turn off the fields $X$ (and $t$ if $m_t^2 \gg m^2$).  In other words, the Higgs mass correction is insensitive to ultraviolet physics. Only the high-frequency scalar mass $m_h$ is small, and the correction (\ref{deltamfin}) is sizable. 

We plot the correction to the Higgs pole mass squared, as a function of the scale $\mu$, in Fig. \ref{f:wilcorr}. We used the Higgs potential relation for the quartic coupling $y_t v =\sqrt{2} m_t, m_h^2 = 2 \lambda v^2$. 

The one-loop correction to the Higgs mass squared by the top quark is calculated perturbatively in Ref. \cite{Choi:2023cqs}. It is also depicted as the dotted curve in Fig. \ref{f:wilcorr}, showing that they match. They are not a priori related because the present effective field calculation has an additional contribution by the high-frequency mode of the Higgs scalar, with the new coupling $\lambda$. 

A similar calculation gives the quartic coupling $\lambda_\mu = \lambda + \delta \lambda(\mu^2)$ from the reference coupling,
\be \label{lambdarun}
\begin{split}
 \delta &\lambda(\mu^2)  =-  9  \int_{\mu^2}^{m_h^2} \frac{dk^2k^2}{(4\pi)^2} \frac{\lambda_k^2}{(k^2+m_k^2)^2}. \\
   &\simeq -\frac{9\lambda^2}{16\pi^2} \left[ \log \frac{2m_h^2}{\mu^2+m_h^2} + \frac{1}{2} - \frac{m_h^2}{\mu^2 + m_h^2}   \right],
\end{split}
 \ee
where we approximated $\lambda_k$ and $m_k$ as before. Note that it is not necessary to match the coupling $\lambda$ at the Higgs pole mass scale as before: we can match $\lambda_k$ at any scale and run from there, and then the mass in (\ref{lambdarun}) is the Higgs mass at the matching scale. 

Besides the well-known logarithmic running, the last term contains information on infrared, where light-scalar correction is essential. The change is up to $0.45 \%$ at 250 GeV, justifying the constancy. 

We have seen that the natural scale parameter for low energy theory is $\mu$.
By differentiating the total mass with the energy scale $\mu$, we obtain the renormalization group equation to one-loop order \cite{Wetterich:1992yh,Morris:1993qb,Morris:1998da}
\begin{align}
  \frac{ dm_\mu^2 }{d\mu^2} \label{massrunning}
 &= - \frac{3\lambda}{32\pi^2} \frac{\mu^2}{\mu^2+m_h^2}\\
 & \quad - \frac{\kappa}{64\pi^2} \frac{\mu^2}{\mu^2+M^2}- \frac{ y_{t}^2\mu^2(\mu^2-m_t^2)}{16\pi^2(\mu^2+m_t^2)^2}, \nonumber\\
\mu \frac{d \lambda_\mu}{d \mu} &= \frac{9 \lambda^2}{8\pi^2 } \frac{ \mu^4 }{(\mu^2+m_h^2)^2}.
\end{align}

Using this, we can study multiply coupled equations by various couplings.
For large $m_h$ and $m_t$, the corresponding fields decouple. For small $m_h$ and $m_t$, the right-hand sides become constants independent of $\mu$, which are commonly used. 

Essentially, the same equations can be obtained by differentiating the bare mass correction \eqref{corrmass} and a similar quartic coupling correction with respect to the upper bound $\Lambda$, instead of the lower bound $\mu$ here. This means that our exact renormalization group equation has a similar form as that in Ref. \cite{Polchinski:1983gv} (focusing on $\hat \phi$)
\be
\begin{split}
\mu \frac {\partial L}{\partial \mu} &=-\mu  \frac{\partial}{\partial \mu} \int_{m_h< |k| <\mu } \frac{d^4k}{(2\pi)^4} \frac{1}{k^2+m_k^2}\\
&\times 
\frac{1}{2}\left[\frac{\partial L}{\partial \hat  \phi(k)} \frac{\partial L}{\partial \hat \phi(-k)}+\frac{\partial^2 L}{\partial\hat  \phi(k) \partial \hat \phi(-k)}\right],
\end{split}
\ee
where $L$ is the momentum-space Lagrangian of the potential density $V$ in (\ref{Masscorr}), using the renormalized couplings, integrated over the same regime $m_h<|k|<\mu$, including the momentum-conserving delta function \cite{Polchinski:1983gv}. It makes the partition function
\begin{align}
{\cal Z} &= \int {\cal D}\phi {\cal D} \hat \phi \exp \left[- \int_{m_h< |k| <\mu } \frac{d^4k}{(2\pi)^4} \right. \nonumber \\
& \left(\textstyle \frac12\hat  \phi(k)(k^2+m_k^2)\hat  \phi(-k) +J(k)\hat \phi(-k) \right) + L \Bigg] \nonumber  \\
&= \int {\cal D}\phi \exp S_\text{eff,$\mu$}
\end{align} 
invariant under the scale change in $\mu$. It defines the Wilsonian effective action $S_\text{eff,$\mu$}$, containing the mass \eqref{mcorrpole} and the quartic interaction (\ref{lambdarun}).  
Conceptually, we track the infrared behavior of the couplings and see at which energy the correcting fields decouple. Also, using the energy-dependent renormalized mass gives a more precise result. 

In conclusion, the observable parameters of a scalar field are governed by light fields only. We also calculated the Higgs mass-squared correction from the pole mass by integrating in its high-energy modes and the top quark.

\begin{acknowledgments}
The author is grateful to Jong-Hyun Baek, Chanju Kim, Hans-Peter Nilles and Piljin Yi for discussions. This work is partly supported by the grant RS-2023-00277184 of the National Research Foundation of Korea.
\end{acknowledgments}

\end{document}